\RequirePackage{fix-cm}
\documentclass[twocolumn,fleqn]{svjour3}          
\smartqed  
\usepackage{graphicx}
\usepackage{mathptmx}      

\usepackage{color}
\usepackage{algorithm}
\usepackage{algorithmicx}
\usepackage{amssymb}
\usepackage{CJK}
\usepackage{url}
\usepackage{ragged2e}
\usepackage{adjustbox}
\usepackage{multirow}
\usepackage[misc]{ifsym}
\usepackage{xcolor}
\usepackage{array}
\usepackage{booktabs}
\usepackage{threeparttable}
\usepackage{algpseudocode}
\usepackage{algorithm,float}
\usepackage[fleqn]{amsmath}
\definecolor{darkgreen}{RGB}{0,180,0}
\floatname{algorithm}{Algorithm}

\begin{document}

\title{A Multi-Domain VNE Algorithm based on Load Balancing in the IoT networks
}
\author{Peiying Zhang\textsuperscript{1}        \and
        Fanglin Liu\textsuperscript{1}        \and
        Chunxiao Jiang(\Letter)\textsuperscript{2}        \and
        Abderrahim Benslimane\textsuperscript{3}        \and
        Juan-Luis Gorricho\textsuperscript{4}       \and
        Joan Serrat-Fern$\acute{a}$ndez\textsuperscript{4}
}
\institute{Peiying Zhang \and Fanglin Liu
            \\\email{25640521@qq.com}
            \\Chunxiao Jiang
            \\\email{jchx@tsinghua.edu.cn}
            \\Abderrahim Benslimane
            \\\email{abderrahim.benslimane@univ-avignon.fr}
            \\Juan-Luis Gorricho
            \\\email{juanluis@entel.upc.edu}
            \\Joan Serrat-Fern
            \\\email{serrat@tsc.upc.edu}
    \\$^{1}$ College of Computer Science and Technology, China University of Petroleum (East China), Qingdao 266580, P.R. China
    \\$^{2}$ Tsinghua Space Center, Tsinghua University, Beijing 100084, P.R. China
    \\$^{3}$ Department of Computer Science, University of Avignon, LIA CERI, Avignon, France
    \\$^{4}$ College of Network Engineering, Technical University of Catalonia (UPC), Carrer de Jordi Girona, Barcelona 1, 3, 08034, Spain
}



\date{Received: date / Accepted: date}

\maketitle

\begin{abstract}
The coordinated development of big data, Internet of Things, cloud computing and other technologies has led to an exponential growth in Internet business. However, the traditional Internet architecture gradually shows a rigid phenomenon due to the binding of the network structure and the hardware. In a high-traffic environment, it has been insufficient to meet people's increasing service quality requirements. Network virtualization is considered to be an effective method to solve the rigidity of the Internet. Among them, virtual network embedding is one of the key problems of network virtualization. Since virtual network mapping is an NP-hard problem, a lot of research has focused on the evolutionary algorithm's masterpiece genetic algorithm. However, the parameter setting in the traditional method is too dependent on experience, and its low flexibility makes it unable to adapt to increasingly complex network environments. In addition, link-mapping strategies that do not consider load balancing can easily cause link blocking in high-traffic environments. In the IoT environment involving medical, disaster relief, life support and other equipment, network performance and stability are particularly important. Therefore, how to provide a more flexible virtual network mapping service in a heterogeneous network environment with large traffic is an urgent problem. Aiming at this problem, a virtual network mapping strategy based on hybrid genetic algorithm is proposed. This strategy uses a dynamically calculated cross-probability and pheromone-based mutation gene selection strategy to improve the flexibility of the algorithm. In addition, a weight update mechanism based on load balancing is introduced to reduce the probability of mapping failure while balancing the load. Simulation results show that the proposed method performs well in a number of performance metrics including mapping average quotation, link load balancing, mapping cost-benefit ratio, acceptance rate and running time.
\end{abstract}

\section{Introduction}
\label{intro}
As representative technologies of the third information revolution, Internet of Things (IoT) \cite{Guo2020UAV}, big data, cloud computing, and edge computing \cite{Zhao2019Computation,Guo2019Toward,Guo2019FiWi} have gradually become an indispensable part of our life through their coordinated development \cite{Du2019Contract,Du2018Auction,DuJ2017AContract}. However, the coordinated development of the three technologies leads to an unprecedented increase in the number and scale of Internet services, the next-generation communication network will face more challenges in the development process, such as dynamic QoS, resource vacancy, network security, and network rigidity. At present, there are many studies on these issues. For example, dynamic spectrum sensing and access technology alleviates the problem of spectrum resource shortage by utilizing spectrum holes \cite{Chunxiao6464633,Chunxiao6489503}, and the non-orthogonal multiple access technology improves resource utilization by multiplexing the power domain and the code domain \cite{Chunxiao7972935}. In addition, the traditional network architecture (that is, every adjustment needs to rebuild the substrate network structure) is also cannot meet the demand, which leads to the problem of Internet rigidity. In this regard, great attention has shifted to network virtualization as a core technology to solve the problem of Internet rigidity \cite{Anderson2005Overcoming,Tutschku2009Network}. The logical networks may transcend substrate infrastructure maintained, and has the advantage of fast configuration, high resource utilization and high isolation capabilities.

The key stage of network virtualization is to map the virtual network (VN) to the substrate network, that is, Virtual Network Embedding (VNE). The VNE problem has been proven to be an NP-hard problem \cite{Amaldi2016On}. Therefore, much work has focused on the research of heuristic algorithms. However, unlike other problems, the components of the solution vector of the VNE problem affect each other, and the order in which different components are solved will affect the solution space of the remaining components. That is, if one of the virtual nodes is mapped to a substrate node first, the other virtual nodes cannot use this substrate node. Therefore, we need to disturb the current solution from time to time in order to get better results, which requires the algorithm to have higher randomness. In addition, the discrete nature of VNE problems may make meta-heuristic algorithms based on direction vectors (such as flower pollination algorithm, differential optimization algorithm, particle swarm algorithm, etc.) invalid. Therefore, the genetic algorithm (GA) based on random search has inherent advantages in solving discrete VNE problems, and has certain optimization value.

Previous work mainly considered the design of algorithm framework, such as: the heuristic algorithm is combined with tabu search algorithm or simulated annealing algorithm to avoid falling into local optimal solution \cite{Yong2007Algorithms,Diallo2019An}, or the mutation operator of GA is added to other heuristic algorithms to increase population diversity. However, the details of the algorithm steps usually retain the traditional design. For example, the crossover probability in the GA is set in a static way, and the mutation gene in the mutation is selected in a random way. This makes the algorithm's running time shorter and the code easier to implement, but the static method is too dependent on experience and cannot flexibly adapt to multiple environments. In addition, when using the Shortest Path algorithm (SP) to estimate the cost of link mapping, the shortest path may not be able to meet the bandwidth resource constraints of the virtual link due to insufficient substrate network resources. However, compared with the traditional network environment, the Internet of Things with a large number of high-demand physical equipment (such as disaster relief, medical, life support equipment) has higher requirements for network stability and algorithm reliability. Therefore, inappropriate fitness estimation methods will result in mapping schemes whose fitness and quality do not match, which will cause a greater impact on the physical world in the IoT environment. In order to solve these problems, we proposed a hybrid GA called LB-HGA based on the traditional GA model.

The main contributions and our main ideas are summarized as follows:

1. In view of the three cases of: both parents' fitness is above average, both parents' fitness is below average, one is better than the mean and the other is worse than the mean, a crossover method based on fitness is proposed. The advantage of this method is that it can not only maintain some randomness, but also effectively the probability of obtaining valid offspring.

2. A mutation gene selection strategy based on pheromone content is proposed. Therein, the pheromone is derived from the ant colony algorithm and is used in this strategy to represent the value of substrate nodes. This strategy can increase or decrease the mutation probability of genes according to their performance. The advantage of this strategy is that it can effectively protect the better offspring obtained by cross operation and improve the probability of the worse offspring being optimized by mutation.

3. A link mapping strategy considering link load balancing and link resource constraints. This strategy can calculate the shortest path that conforms to different resource constraints, which can make the link cost estimation more accurate in the fitness calculation.

The reminder of this paper is organized as follows. Section 2 reviews the existing methods for VNE. Section 3 introduces the network model and problem statement. Section 4 introduces the three core strategies used in LB-HGA method. In Section 5, we describe our proposed method LB-HGA in detail. The performance of our method and other methods is evaluated in Section 6. Section 7 concludes this paper.
\section{Related Work}
\label{sec:1}
A classification strategy \cite{Cao2018Heuristic} based on algorithm logic divides existing VNE methods into optimal algorithm and heuristic algorithm in which the heuristic algorithms can be further divided into traditional heuristic algorithm and meta-heuristic algorithm. Whereas the solution obtained by the optimal algorithm is closer to the optimal solution, these are characterized by high computational time which renders unsuitable for practical delay sensitive scenarios. On the other hand, heuristic algorithms often cannot guarantee an optimal solution but have an appealing characteristic of low time complexity. Therefore, the two approaches present a trade-off between solution Optimality and execution time.
\subsection{Optimal Algorithms}
\label{sec:2}
A typical optimization algorithm is proposed in \cite{Lischka2009A} in which the authors proposed a VNE algorithm based on subgraph isomorphism detection. This method has a good mapping effect for large. In the same year, the authors of \cite{Chowdhury2009Virtual} for the first time applied a mixed integer linear programming(MIP) model to solve the VNE problem and proposed D-ViNE and RViNE algorithms based on LP relaxation to tame the time complexity of the MIP algorithm. However, this work has less coordination between the two mapping phases (link mapping and node mapping). In order to make up this defect, the authors of \cite{Gao2010A} proposed a progressive greedy VNE algorithm (PG-VNE), which is shown to result into better coordination between the two phases. In addition, with the development of IoT and other technologies to improve the demand for network service quality, the authors of \cite{articleC} proposed a dynamic mapping algorithm based on QoS driver to further meet the demands of customized QOS. In the following year, the authors of \cite{articleFF} further considers the perception of energy consumption, avoiding the single consideration of mapping revenue. In recent studies, the authors of \cite{articleHH} proposed a candidate set based mapping algorithm considering delay and geographical location constraints, which is significantly less time complexity than the existing heuristic algorithms. In view of the lack of multi-attribute decision making in the single-domain mapping problem, the authors of \cite{articleAAA} proposed a new node ordering method, which comprehensively considered five topology attributes and global network resources, and showed good performance.

Mathematically speaking, any optimization method involves finding the extremum under certain constraints. But in the case of a larger problem which is the case in most scenarios, solving the optimal solution tends to consume large amounts of computing resources. For this reason, the optimal method in the large-scale network environment is not widely used. Therefore, the study of heuristic algorithm which gives a feasible solution at an acceptable cost is important.
\subsection{Heuristic Algorithms}
\label{sec:3}
In the classical algorithm \cite{Yong2007Algorithms}, a greedy algorithm based on node matching is used for node mapping, and k-shortest path is used for link mapping. In addition, the authors of \cite{Zhang2013A} proposed a unified enhanced VN embedding algorithm (VNE-UEPSO) based on particle swarm optimization (PSO). However, the algorithm has higher randomness and slower convergence speed. In order to overcome this commonly occurring shortcoming, the authors of \cite{Li2014Virtual} proposed a PSO optimization scheme based on probabilistic particle reconstruction. The algorithm sacrifices some computation time, but the result is better than the traditional PSO algorithm. In addition to the PSO algorithm, GA has also attracted wide attention because of its excellent performance. The authors of \cite{Xiuming2012An} proposed a VNE strategy (CB-GA) based on the simple node sorting method and GA. The authors of \cite{Pathak2017A} proposed a GA model based on new chromosomes to solve the multi-domain VNE problem. However, both of these algorithms rely on probability for random selection, crossover and variation, so it is difficult to guarantee that an excellent enough solution can be found within a limited number of iterations. In order to make up for these shortcomings, in recent studies, the authors of \cite{Zhuang2018A} proposed a virtual network mapping strategy based on cellular automata genetic mechanism. The algorithm introduced cellular automata model of network nodes, effectively guides the crossover stage, ensures the diversity of population, and avoids premature convergence. However, since the mutation operation of this algorithm has random variation, the unguided random variation may cause the better individuals that were selected to mutate into the worse ones. Moreover, the algorithm does not clearly consider the load balancing of nodes and links, so there is still some room for optimization.

Based on the above analysis, it can be seen that as far as genetic algorithms are concerned, there is still some room for optimization in the current research.
\section{Network Model and Problem Statement}
\begin{figure}
\center{\includegraphics[width=8.5cm]  {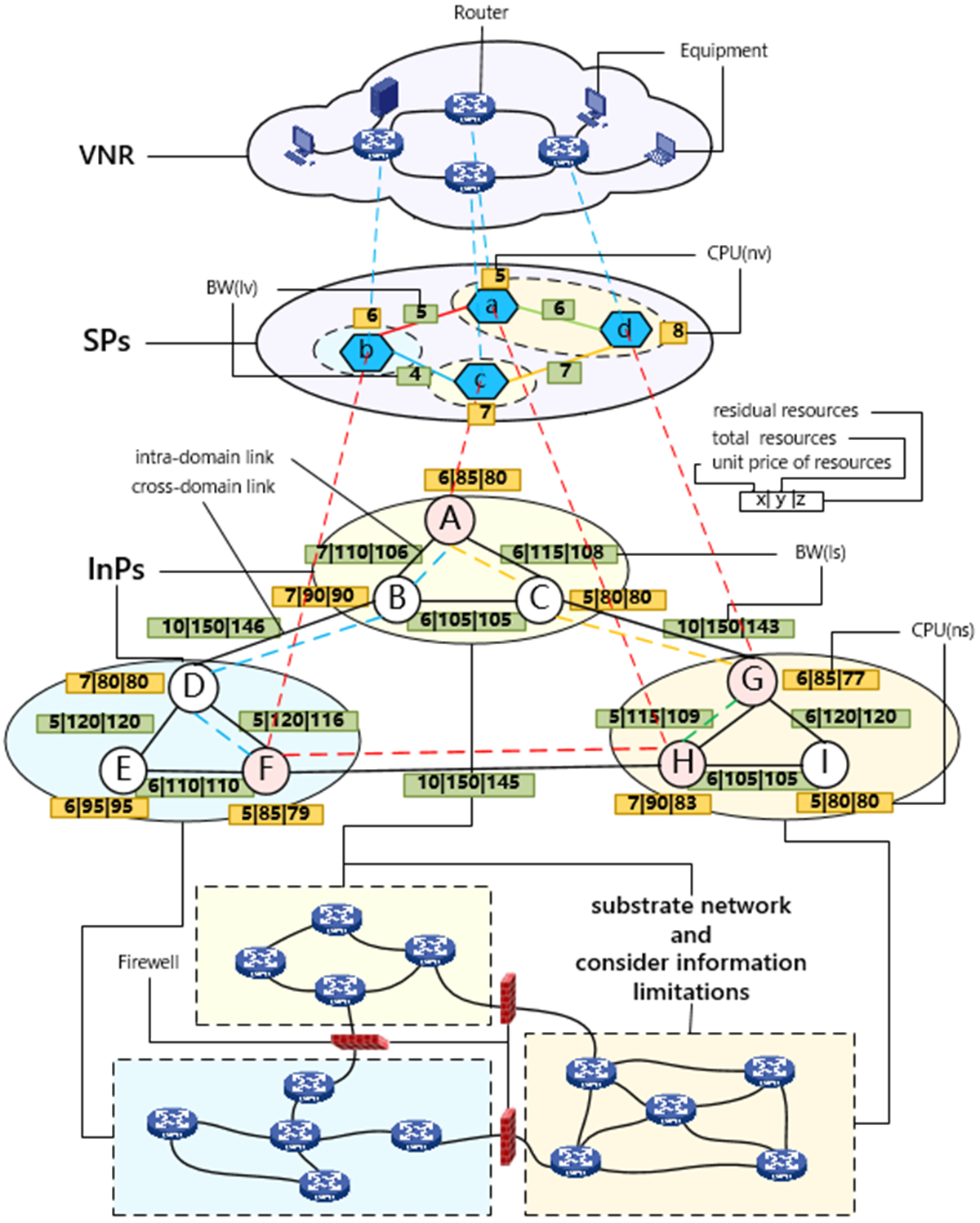}}
\caption{The mapping process of the virtual network to the substrate network.}
\label{fig:1}
\end{figure}
\label{sec:4}
\subsection{Substrate Network and Virtual Network Model}
Figure.~\ref{fig:1} shows the mapping process consisting of four layers, and the tags on the picture apply to the full-text picture. The substrate network is abstracted as an undirected graph $G^s = \{N^s,L^s\}$, where $N^s$ represents the set of substrate nodes and $L^s$ represents the set of substrate links. Each substrate node has functional or non-functional attributes, including the CPU capacity $CPU(n_s)$, and the unit price $UP(n_s)$ of $CPU(n_s)$. Each substrate link also has a set of attributes, including the bandwidth $BW(l_s)$ and the unit price $UP(l_s)$ of $BW(l_s)$. We define the set of substrate paths as $P_s$. And a substrate path set from substrate node $i$ to substrate node $j$ is represented by $P_s(i,j)$. Similarly, a VN can also be abstracted as a weighted undirected graph $G^v = \{N^v,L^v\}$, and in each Virtual Network Request (VNR), $N^v$ represents the set of virtual nodes and $L^v$ represents the set of virtual links. Each virtual node $n_v\in N^v$ has a requirement for CPU, that can be defined as $CPU(n_v)$. And each virtual link $l_v\in L^v$ has a requirement for bandwidth, that can be defined as $BW(l_v)$.
\subsection{Virtual Network Embedding Problem Description}
The process can be modeled as a mapping M: $G^v\{N^v, L^v\}\to G^s\{N^s, P_s\}$. The VNR mapping process consists of two steps: (i) virtual node mapping; (ii) virtual link mapping;. In the node mapping phase, each virtual node $n_v\in N^v$ chooses a substrate node that conforms to the constraint condition as the mapping target. Different virtual nodes in the same VNR cannot be mapped repeatedly to the same substrate node. In the link mapping phase, each virtual link $l_v\in L^v$ in the VN is mapped to an substrate path $P_s(l_v)$.
\subsection{Objectives and Evaluation Index}
Since the cost of mapping nodes is certain, some studies omit it in the objective function and only retain the cost of bandwidths. However, since we consider that different domains in the multi-domain substrate network have the different unit prices of CPU, so our objective function will consider the cost of CPU. Model it as an integer programming model and shown below:
\begin{equation}
\begin{aligned}
OBJ(VN) = &min\sum_{n_v\in N^v}CPU(n_v)\times UP(n_s)+\\
&\sum_{l_v\in L^v}BW(l_v)\times AUP(P_s),
\end{aligned}
\label{eq:1}
\end{equation}
\begin{equation}
AUP(P_s)=\sum_{l_s\in P_s}UP(l_s),
\label{eq:2}
\end{equation}
where $AUP(P_s)$ represents the aggregate unit price of path $P_s$.

In addition, the mapping needs to meet the constraints of VNR. In this model, it can be formulated as:
\begin{equation}
\begin{cases}
BW(l_v)\leq BW(l_s),\;\forall l_s\in Ps(l_v),\\
CPU(n_v)\leq CPU(n_s),\;n_v\leftrightarrow n_s,
\end{cases}
\label{eq:3}
\end{equation}
where $\leftrightarrow$ represents the two ends of the arrow map to each other.

We use 5 evaluation indexes to measure the performance of VNE algorithms. Including the load balancing of substrate links, the ratio of revenue to cost, the VN request acceptance ratio, the mapping average quotation, and the running time of algorithms. Therein, the running time of algorithms includes the average running time and the total running time. In addition, and we use the mapping average earnings to assist the illustration.

We use the variance of bandwidths' consumption to measure the link load balancing, and it can be formulated as follows:
\begin{equation}
\sigma^2=\frac{\sum_{l_s\in L^s}(BC(l_s)-\mu)}{N},
\label{eq:4}
\end{equation}
where $BC(l_s)$ represents the consumption of bandwidths of the substrate link $l_s$, it can be formulated as total $BW(l_s)$ - residual $BW(l_s)$. $\mu$ represents the population mean of $BC(l_s)$, and $N$ is the number of links in the substrate network.

The revenue of mapping a VN at time t can be defined as the resources for all virtual nodes and virtual links requested by the VN, and it can be formulated as follows:
\begin{equation}
R(G^v,t)=\sum_{n_v\in N^v}CPU(n_v)+\sum_{l_v\in L^v}BW(l_v).
\label{eq:5}
\end{equation}

The cost of mapping a VN at time $t$ can be defined as the total amount of substrate network resources that allocated to the VN, and it can be formulated as follows:
\begin{equation}
C(G^v,t)=\sum_{n_v\in N^v}CPU(n_v)+\sum_{l_v\in L^v}BW(l_v)Hops(P_s(l_v)),
\label{eq:6}
\end{equation}
where $Hops(P_s(l_v))$ represents the number of hops of the substrate path $P_s(l_v)$ that the virtual link $l_v$ eventually mapped to.

Based on the above model, the revenue to cost ratio over a period of time $t\in(0,k)$ can be formulated as follows:
\begin{equation}
\frac{R}{C}=\frac{\sum_{t=0}^kR(G^v)}{\sum_{t=0}^kC(G^v)}.
\label{eq:7}
\end{equation}

The VN request acceptance ratio over a period of time $t\in(0,k)$ can be defined as follows:
\begin{equation}
acceptance\;ratio=\frac{\sum_{t=0}^kVNR_{accept}}{\sum_{t=0}^kVNR_{refuse}},
\label{eq:8}
\end{equation}
where $VNR_{accept}$ represents the number of VNRs that were accepted and successfully mapped, and $VNR_{refuse}$ represents the number of rejections.

The mapping quotation is defined as the price the user has to pay to map a VN, it's the same as Equation~\ref{eq:1}. The average quotation is the average price of mapping VNRs over a period of time $t\in(0,k)$, and it can be formulated as follows:
\begin{equation}
average\;quotation=\frac{\sum_{t=0}^kOBJ(VN)}{\sum_{t=0}^kVNR_{accept}}.
\label{eq:9}
\end{equation}

The total running time is the total time that each algorithm runs in a simulation experiment, and the time is measured in milliseconds. In addition, the average running time can be formulated as follows:
\begin{equation}
average\;time=\frac{total\;time}{\sum_{t=0}^kVNR_{accept}},
\label{eq:10}
\end{equation}
\section{Strategy Model and Innovation Motivations}
In this section, we introduce the core strategies used in LB-HGA algorithm in detail. We will analyze the problems existing in traditional algorithms, give the motivations of optimization, and give the required mathematical expression. In addition, these strategies will be used in the next section as part of the algorithm model.
\subsection{Dynamic Crossover Probability}
The crossover probability in traditional GA models is mostly fixed, such as literatures \cite{JiangRenewal,Cai2013A,Zhou2014Virtual,Pathak2017A,JiangJoint}. This makes the algorithms computational complexity small and the code implementation simple. But it will make the parents with different performance have the same crossover probability. However, the upside potential of different individuals is different (which is usually related to the fitness of the individuals). We believe that different crossover probabilities should be calculated for different quality parents in order to improve the possibility of obtaining excellent offspring.
\begin{figure}[htbp]
\small
\centering
\includegraphics[width=8cm]{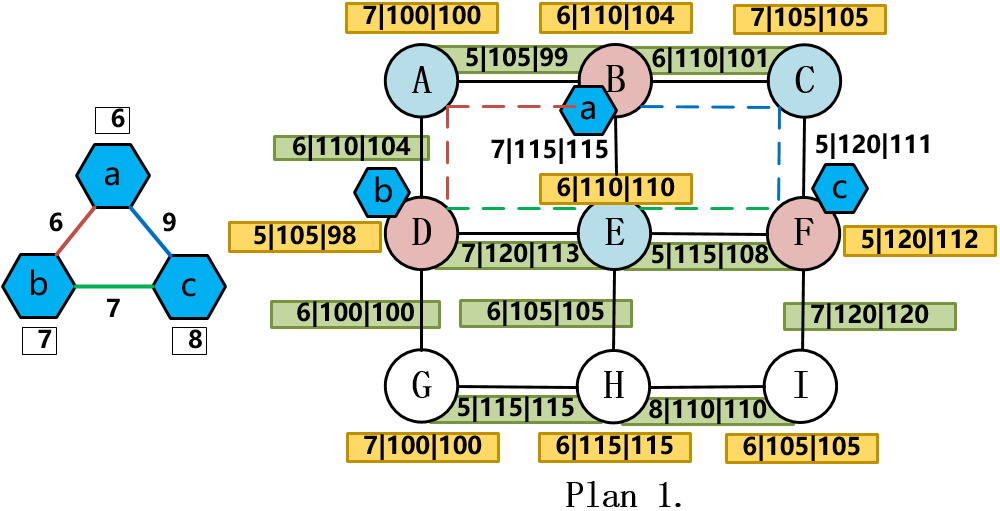}\\
\caption{The diagram of a VNR and mapping plan 1 with fitness of 53.}
\label{fig:2}
\end{figure}
\begin{figure}[htbp]
\small
\centering
\includegraphics[width=8cm]{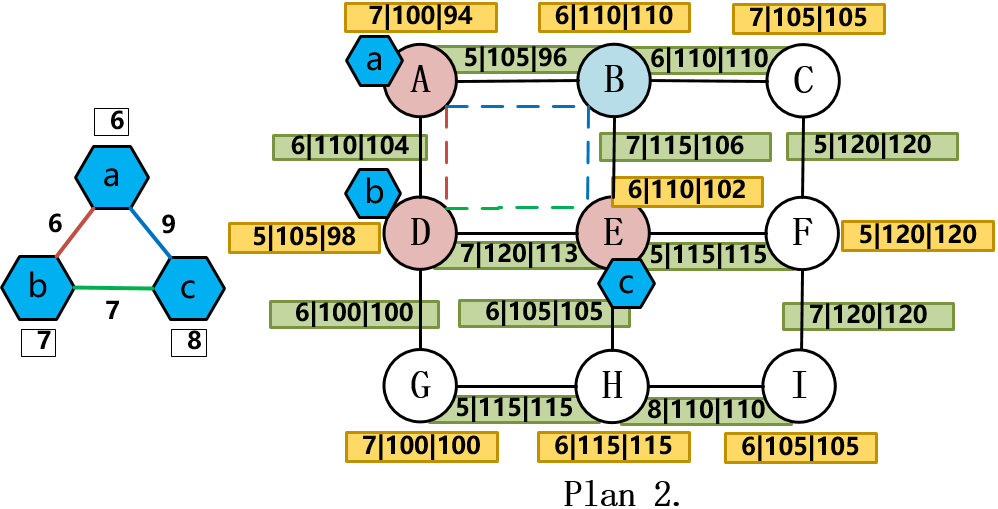}\\
\caption{The diagram of a VNR and mapping plan 2 with fitness of 43.}
\label{fig:3}
\end{figure}
\begin{figure}[htbp]
\small
\centering
\includegraphics[width=8cm]{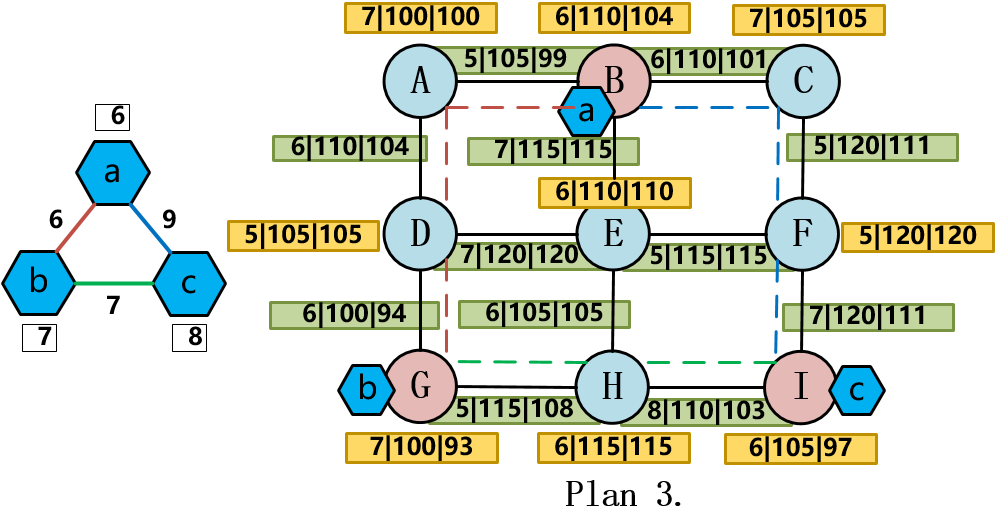}\\
\caption{The diagram of a VNR and mapping plan 3 with fitness of 67.}
\label{fig:4}
\end{figure}
\begin{figure}[htbp]
\small
\centering
\includegraphics[width=8cm]{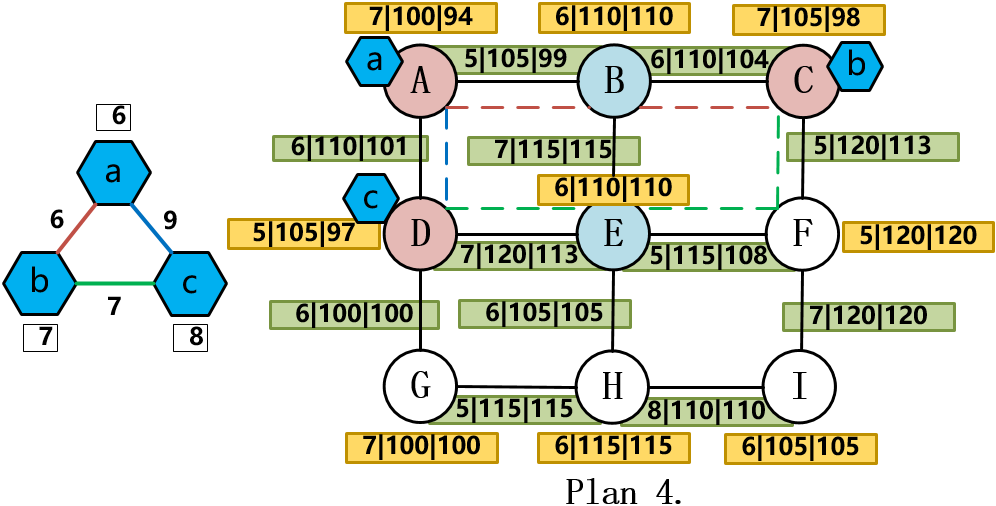}\\
\caption{The diagram of a VNR and mapping plan 4 with fitness of 51.}
\label{fig:5}
\end{figure}

As illustrated in Fig. ~\ref{fig:2}-~\ref{fig:5}, on the left is an example of a VNR, and on the right is a solution for mapping this VNR. Taking virtual node C as an example, the better choices (BCs) in each plan are marked blue. Therein, BCs mean the alternative mapable substrate nodes that the virtual nodes can choose to make the fitness lower. As can be seen from Fig.~\ref{fig:4} and Fig.~\ref{fig:3}, plan 3 with the highest fitness has 6 BCs, while plan 2 with the lowest fitness has 1 BCs. Thus, it can be seen that the plan with better performance has smaller ascending space than the plan with poorer performance. In addition, although BCs can more accurately reflect the upside potential, calculating the number of BCs for each parents will make the calculation too much. In order to balance the running time and performance, we designed the following crossover probability function based on fitness.

1. $min\{F(x_1),F(x_2)\}$ $\geq\;\bar{X}$:
\begin{equation}
P(x_1,x_2)=\frac{\lambda_1\times(min\{F(x_1),F(x_2)\}-\bar{X})}{max\{F(x_1),...,F(x_n)\}-\bar{X}}.
\label{eq:11}
\end{equation}
\begin{equation}
\bar{X}=\frac{F(x_1)+F(x_2)+,...,+F(x_n)}{n},
\label{eq:12}
\end{equation}
where $F(x_i)$ represents the fitness of the individual $x_i$, and $\lambda_1$ intervenes in the crossover probability with the default value of 1 and the adjustment range of (0,2].

2. $max\{F(x_1),F(x_2)\}$ $\leq\;\bar{X}$:
\begin{equation}
P(x_1,x_2)=\frac{\lambda_2\times(1-(\bar{X}-max\{F(x_1),F(x_2)\}))}{\bar{X}-min\{F(x_1),...,F(x_n)\}},
\label{eq:13}
\end{equation}
where the default value and range of $\lambda_2$ are the same as $\lambda_1$. And the $\lambda_2$ is recommended to set $\lambda_2$ to the default value or slightly smaller than 1.

3. $min\{F(x_1),F(x_2)\}$ $< \bar{X}$ and $max\{F(x_1),F(x_2)\}$ $> \bar{X}$:
\begin{equation}
S_{max}=\frac{max\{F(x_1),F(x_2)\}-\bar{X}}{max\{F(x_1),...,F(x_n)\}-\bar{X}},
\label{eq:14}
\end{equation}
\begin{equation}
S_{min}=\frac{\bar{X}-min\{F(x_1),F(x_2)\}}{\bar{X}-min\{F(x_1),...,F(x_n)\}},
\label{eq:15}
\end{equation}
\begin{equation}
P(x_1,x_2)=\left\{
\begin{array}{rcl}
\lambda_1\times S_{max} & & {S_{max} > S_{min},}\\
\lambda_2\times (1-S_{min})& & {S_{max} \leq S_{min}.}\\
\end{array} \right.
\label{eq:16}
\end{equation}

In the third case, the fitness of the parents is better or worse than the overall average fitness of the population, respectively. Therefore, further analysis is needed to identify individuals in parents who deserve more attention. $S_{max}$ represents the importance of the individual with high fitness. $S_{min}$ represents the importance of the individual with low fitness. Function~\ref{eq:15} means that the crossover probability will consider the more important individual and multiply the corresponding intervention weight according to the tendency to support or oppose crossover.
\subsection{Link Load Balancing Strategy}
\begin{figure}[htbp]
\small
\centering
\includegraphics[width=8cm]{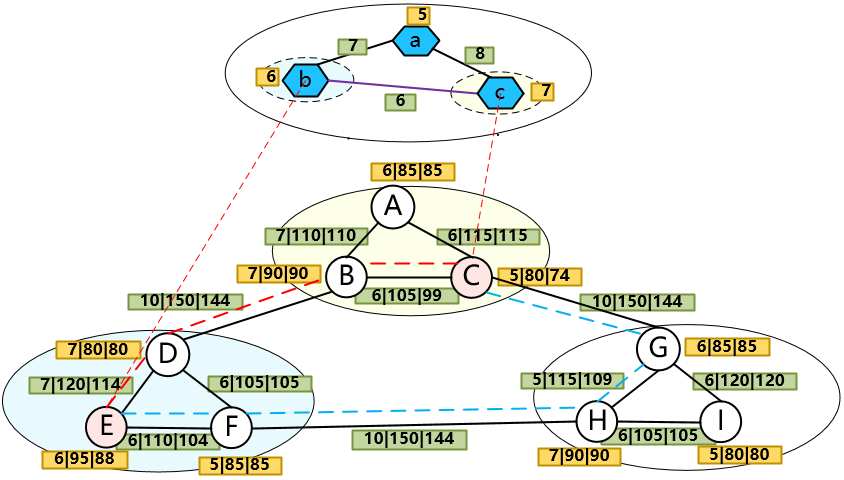}\\
\caption{The diagram of a multidomain substrate network with the initial weights.}
\label{fig:6}
\end{figure}

The static weights that does not take the load balancing into account will cause the resources of the substrate links with less weighted decrease too fast. And when the substrate network resources are relatively small, the SP algorithms that do not take resource constraints into account may not be able to obtain the mapping scheme of links conforming to the constraints. This will make the estimation of individuals' fitness in the node mapping stage inaccurate, as shown in Fig.~\ref{fig:6}.

Figure.~\ref{fig:6} shows a substrate network with three physical domains and a VNR. In addition, the virtual link $l_s(b,c)$ in the VN is mapped to the substrate path $P_s(E,C)$. When we set the weight as $UP(l_s)$, the $P_s(E,C)\{E\to D\to B\to C\}$ is the shortest path when resources are abundant, and the $P_s(E,C)\{E\to F\to H\to G\to C\}$ is the shortest path when the resources of $P_s(E,C)\{E\to D\to B\to C\}$ are scarce. The link aggregation unit price difference between the two is 10, the difference is large. If load balancing is not considered, the substrate network resources will uneven occupancy in the later stage of mapping, and some paths will get blocked, which will lead to the increase of response time and the increase of mapping cost. However, if load balancing is considered, the VNRs later can also get a better mapping scheme.

A simple way to consider load balancing is to adjust the weight of the substrate link according to the bandwidth occupancy of the substrate link. It can be formulated as:
\begin{equation}
W(l_s)=\left\{
\begin{array}{rcl}
UP(l_s)(1+\lambda\times extra\;weight) & & {U(l_s) > \bar{U},}\\
UP(l_s)& & {U(l_s) \leq \bar{U}.}\\
\end{array} \right.
\label{eq:17}
\end{equation}
\begin{equation}
extra\;weight=\frac{U(l_s)-\bar{U}}{max\{\forall l_s\in L_s\vert U(l_s)\}-\bar{U}},
\label{eq:18}
\end{equation}
\begin{equation}
\bar{U}=\frac{\sum_{l_s\in L_s}U(l_s)}{n},
\label{eq:19}
\end{equation}
\begin{equation}
U(l_s)=\sum_{l_v\in M(l_s)}BW(l_v),
\label{eq:20}
\end{equation}
where the range of $\lambda$ is (0,2], $\bar{U}$ represents the average used bandwidth of n substrate links in substrate network, $U(l_s)$ represents the total amount of bandwidths used in a substrate link $l_s$, and $M(l_s)$ represents a collection of mapped virtual links on a substrate link $l_s$. Equation~\ref{eq:16} means that when the used bandwidth $U(l_s)$ of a substrate link is larger than the average used bandwidth $\bar{U}$ of substrate network, the weight will increase with the increase of $U(l_s)$. When $U(l_s)$ is less than $\bar{U}$, then use $UP(l_s)$ as link weight. By adding intervention weight $\lambda$, the manager can adjust the importance of load balance according to the demand, and make the algorithm more flexible.

\begin{figure}[htb]
 \center{\includegraphics[width=8cm]  {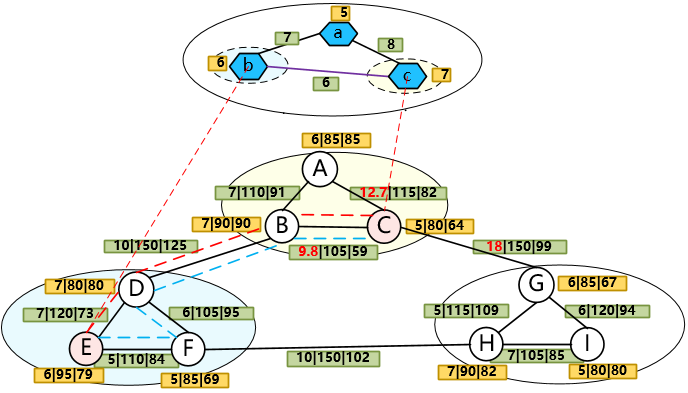}}
 \caption{The diagram of a multidomain substrate network that consumes a portion of bandwidth resources with the weight that considering the load balance.}
 \label{fig:7}
\end{figure}

Some bandwidth resources in the substrate network as shown in Fig.~\ref{fig:6} are randomly consumed to form the substrate network as shown in Fig.~\ref{fig:7}.
The intervention weight $\lambda$ was set to 0.8, and the weight of all links in the substrate network was adjusted. After adjustment, the weight with changes was marked as red. The $P_s(E,C)\{E\to D\to B\to C\}$ is the shortest path before weight adjustment and the $P_s(E,C)\{E\to F\to D\to B\to C\}$ is the shortest path after weight adjustment. It can be seen that after weight adjustment, the mapping can bypass the links with high consumption of bandwidth resources.

In the stage of GA, single source shortest path is suitable for the algorithms with both paths and nodes in individuals. Since the $BW(l_v)$ required by each virtual link $l_v\in L^v$ is not the same, the shortest path needs to be calculated for different links. The multi-source shortest path is suitable for the algorithms that only includes nodes in individuals. Because the multi-source SP algorithms is only used to estimate the cost of mapping of virtual links when calculating fitness, it is not necessary to consider the exact resource constraints. Moreover, after the node mapping stage, the mapping scheme of virtual links needs to be obtained by using an single source SP algorithm. At this time, the precise resource constraints need to be considered. In addition, when solving the single source shortest path, the bandwidth resources required by each virtual link can be taken as the constraint. By setting the weight of the bottom link with insufficient resources to be the highest, it can be prevented from being selected into the mapping scheme, thereby preventing mapping failure. When solving the multi-source shortest path, only the minimum resource constraints needs to be satisfied. And the minimum resource constraints is equal to the $BW(l_v)$ of the virtual link $l_v$ that requires the least bandwidth resources in the unmapped VN.
\subsection{Gene Selection Strategy}
\newcommand{\tabincell}[2]{\begin{tabular}{@{}#1@{}}#2\end{tabular}}
We consider a gene selection strategy to introduce the concept of pheromones in Ant Colony Algorithm (ACA) into GA to guide the selection of mutation nodes. The introduction of ACA can be obtained from \cite{Dorigo2005Ant}, and there are some examples of genetic algorithms being combined with ant colony algorithms in literatures \cite{Shang2007Hybrid,Ming2018Network,Wei2011Ant}. In one iteration, individuals with lower fitness will release more pheromones, and individuals with higher fitness will release fewer pheromones. In the mutation stage, the nodes with lower pheromones will be more likely to selected for mutation. Introducing the positive feedback mechanism into the genetic algorithms will increase the interactivity of the population and reasonably guide the selection of mutation nodes.

In addition, we provide a pheromones initialization strategy for the initial population, and it can be abstracted as the following function:
\begin{equation}
\tau_{ns}(t)=\sum_{k=1}^{num(X)}\Delta(1)\tau_{ns}^k, n_s\in N^s,
\label{eq:21}
\end{equation}
\begin{equation}
\Delta(1)\tau_{ns}^k=\left\{
\begin{array}{rcl}
\frac{max\{F(x_i),x_i\in X\}-F(x_k)}{num(N_k^s)}& & {n_s\in x_k,}\\
0& & {n_s\not\in x_k,}\\
\end{array} \right.
\label{eq:22}
\end{equation}
where $\tau_{ns}(t)$ represents the pheromones quantity of the substrate node $n_s$ when the number of iterations is t, $num(X)$ represents the number of individuals in the population $X$, $num(N_k^s)$ represents the number of substrate nodes of the individual $x_k$, and $\Delta(1)\tau_{ns}^k$ represents the pheromones released by the individual $x_k$ on the substrate node $n_s$.

The pheromone update strategy of the crossover stage can be abstracted as the following function:
\begin{equation}
\tau_{ns}(t+1)=(1-\rho)\tau_{ns}(t)+\sum_{k=1}^{num(X)}\Delta(1)\tau_{ns}^k, n_s\in N^s,
\label{eq:23}
\end{equation}
where $\rho$ represents the pheromones dissipation factor. In addition, Equation~\ref{eq:19} indicates that after reducing pheromones in a certain proportion, all the new individuals generated by crossover in one iteration will leave pheromones in the substrate nodes of individuals according to their fitness. Moreover, since the goal of our algorithm is to minimize fitness, we take the difference between the fitness of each individual in the population and the highest fitness in the population as the reference for pheromone updates to reflect the goal.

During the mutation state, the pheromone update rules for each node in the individual $x_i$ be selected for mutation can be abstracted as the following function:
\begin{equation}
\tau_{ns}(t+1)=\left\{
\begin{aligned}
\tau_{ns}(t)-\Delta(2)\tau_{ns}^i& & {F(x_i)^{before}>F(x_i)^{after},}\\
\tau_{ns}(t)& & {F(x_i)^{before}=F(x_i)^{after},}\\
\tau_{ns}(t)+\Delta(2)\tau_{ns}^i& & {F(x_i)^{before}<F(x_i)^{after},}\\
\end{aligned}
\right.
\label{eq:24}
\end{equation}
where $n_s\in mutant\;gene\;set$, $mutant\;gene\;set$ is defined as a set of genes selected for mutation in $x_i$, $F(x_i)^before$ represents the fitness of the $x_i$ before mutation, and $F(x_i)^after$ represents the fitness of $x_i$ after mutation.
\begin{equation}
\tau_{ns}(t+1)=\left\{
\begin{aligned}
\tau_{ns}(t)+\Delta(2)\tau_{ns}^i& & {F(x_i)^{before}>F(x_i)^{after},}\\
\tau_{ns}(t)& & {F(x_i)^{before}=F(x_i)^{after},}\\
\tau_{ns}(t)-\Delta(2)\tau_{ns}^i& & {F(x_i)^{before}<F(x_i)^{after},}\\
\end{aligned}
\right.
\label{eq:25}
\end{equation}
where $n_s\in goal\;node\;set$, the $goal\;node\;set$ is defined as a set of goal nodes that selected by $mutant\;gene\;set$, and it can also be called post-mutation nodes.

The $\Delta(2)\tau_{ns}$ of the mutation stage is different from the $\Delta(1)\tau_{ns}$ of the crossover stage, and it can be formulated as:
\begin{equation}
\Delta(2)\tau_{ns}^i=\frac{|F(x_i)^{after}-F(x_i)^{before}|}{num(mutant\;gene\;set)},
\label{eq:26}
\end{equation}
where $\Delta(2)\tau_{ns}^i$ represents the pheromones released by the $x_i$ on the substrate node $n_s$, and $num(mutant$ $gene\;set)$ represents the number of genes in $mutant\;gene\;set$.

According to the proportion of pheromones amount of each node in the mutation stage to the total pheromones amount of all substrate nodes in the individual, a certain number of different mutation genes were obtained by roulette algorithm, and these genes were used to form $mutant\;gene\;set$ for mutation. Where, the proportion of pheromones can be formulated as follows:
\begin{equation}
pheromone\;proportion=\frac{\tau_{ns}(t)}{\sum_{n_s\in X_i}\tau_{ns}(t)}.
\label{eq:27}
\end{equation}

In addition, because all the substrate nodes of the individual must be released pheromones in the crossover stage, so $\tau_{ns}(t)$ must be greater than 0.
\section{Heuristic Algorithm Design}
Based on the dynamic crossover probability, the load balancing and the resource constraints strategy, and the gene selection strategy, a hybrid GA for VNE problem solving strategy LB-HGA is proposed.
\subsection{Node Mapping Algorithm}
We use the optimized GA to complete the mapping of nodes. In this model, we take the real number encoding method and define the individuals as $X_i=\{X_i^1,X_i^2,...X_i^j...X_i^n\}$, where $X_i$ represents the individual numbered $i$ in the population. In addition, $n$ is the number of virtual nodes in the virtual network, $x_i^j$ represents the substrate node corresponding to the virtual node numbered $j$, and the gene belongs to the individual $X_i$. And we use Equation~\ref{eq:1} as the fitness function $F(x_i)$.

We modified the iterative steps based on the framework of the traditional GA algorithm. Therein, the elite selection strategy was adopted to retain half of the individuals with lower fitness. For cross process, select a pair of individuals at random and decide whether to generate offspring through the dynamically calculated crossover probability. If crossover is determined, several pairs of alleles are randomly selected and exchanged. In addition, for each newly generated offspring, mutation is determined according to a certain probability. Moreover, a strategy named cataclysm is used to jump out of the local optimal solution. It occurs when the maximum number of iterations $\times$ 0.6 consecutive iterations do not update the optimal solution. Only the first third of the individuals with the lowest fitness were retained, and then the initialized individuals were generated to complete the population, so that the number of individuals in the population was maintained at $X$.

The detailed steps of node mapping algorithm are illustrated in Algorithm 1.
    \begin{algorithm}
        \caption{The Node Mapping Algorithm Based on Optimized Genetic Algorithm.}
        \begin{algorithmic}[1] 
            \Require Substrate network $G^s = \{N^s,L^s\}$, and virtual network request $G^v = \{N^v,L^v\}$.
            \Ensure Virtual network node mapping scheme.
            \State $P_m$ $\gets$ mutation probability;
            \State X $\gets$ maximum population capacity;
            \State k $\gets$ 0;
            \State $P_c \gets$ 0;
            \State Randomly generate X individuals;
            \State Initialize the pheromones in the substrate network;
                \For{not reached max iterations}
                    \State Use elite choice strategy to select $\frac{X}{2}$ individuals;
                    \While{the number of individuals is less than X}
                        \State Select a pair of individual at random and calculate the cor-
                        \State responding crossover probability $P_c$;
                        \If {random decimal $ < P_c$}
                            \State Crossing this two individuals;
                        \EndIf
                        \State Feasibility judgment;
                    \EndWhile
                    \State Update the pheromones based on pheromone update strategy;
                    \For{each individual in the new offspring}
                        \State Calculate fitness;
                        \If {random decimal $ < P_m$}
                            \State Mutation;
                        \EndIf
                    \EndFor
                    \State Update the pheromones of genes in the $mutant\;gene$
                    \State $set$ and $goal\;node\;set$;
                    \If {the optimal solution has been updated}
                        \State k = 0;
                    \Else
                        \State k++;
                        \If {k == max iterations $\times$ 0.6}
                            \State Cataclysm;
                            \State k = 0;
                        \EndIf
                    \EndIf
              \EndFor
            \State \Return{The individual with the lowest fitness;}
        \end{algorithmic}
    \end{algorithm}
\begin{algorithm}
        \caption{The Link Mapping Algorithm Based on Shortest Path Algorithm.}
        \begin{algorithmic}[1] 
            \Require Virtual network node mapping scheme.
            \Ensure Virtual network link mapping scheme.
            \State Sort the virtual links by the required bandwidth in nonincreasing order;
            \For{all the unmapped virtual links in VNR}
                \State Gets the corresponding two substrate endpoints;
                \State Update the weight of each substrate link;
                \State Obtain the shortest path between the two endpoints;
            \EndFor
            \State \Return{Link mapping scheme;}
        \end{algorithmic}
    \end{algorithm}
\subsection{Link Mapping Algorithm}
The detailed steps of link mapping algorithm are illustrated in Algorithm 2.
\section{Performance Evaluation}
In this section, we describe the setup of the simulation environment, including the parameters of the substrate network and algorithm, and give the experimental results. We used the five evaluation criteria defined earlier to measure the performance of our method against the others. In addition, we also describe the mapping process and parameter setting of other algorithms.
\subsection{Environment Settings}
The experiment was run on a PC with Intel Core i5 2.90GHz CPU and 8 GB memory. The substrate network topology and virtual network request topology are generated by the GT-ITM \cite{Zegura1996How} tool. The substrate network includes a total of 4 domains, and each domain includes 30 substrate nodes. Therein, the CPU capacity of the substrate nodes ranges from [100,300], the bandwidth of the links within the domain ranges from [1000,3000], and the bandwidth of the inter-domain links ranges from [3000,6000]. The unit price of the bandwidth and the unit price of the CPU are both in the range of [1,10]. In addition, the value range of the number of virtual nodes in a VN is [5,10], and the value range of the CPU capacity required by the virtual node and the bandwidth resource required by the virtual link are both [1,10]. The above variables all obey uniform distribution. In addition, the number of VNRs follows a Poisson distribution with an average of 10 within 100 time units. The simulation time is 2200 time units, and the life of the VN is 1000 time units.
\subsection{Algorithm Parameters}
We compared the designed algorithm with the other three existing heuristic VNE problem solving methods. Table 1 shows the comparison and introduction of the mapping process of the other three algorithms, and Table 2 shows the parameter settings of the all four algorithms.
\begin{table}[htbp]
\caption{Introduction of three algorithms compared with LB-HGA.}
\label{tab:3}
\begin{tabular}{|p{0.9cm}|p{6.6cm}|}
\hline
Notation&Description\\
\hline
T-GA&Traditional GA and SP algorithm are used.\\
\hline
MDPSO&For each virtual node in VN, obtain a group of candidate nodes and then use particle swarm optimization algorithm to obtain the mapping scheme of virtual nodes. The mapping of virtual links is based on the traditional SP algorithm.\\
\hline
IVERM&The single domain mapping is carried out when there are enough resources in alternative domain, the cross-domain mapping based on genetic algorithm is used when resources are insufficient. The mapping of virtual links is based on the shortest path algorithm.\\
\hline
\end{tabular}
\end{table}
\begin{table}[htbp]
\caption{Parameter setting of four algorithms.}
\label{tab:4}
\begin{tabular}{|p{1.2cm}|p{6.3cm}|}
\hline
MDPSO&The number of particles and iterations are 10 and 50, and the $\omega$, $\rho$1 and $\rho$2 in the velocity update formula are 0.1, 0.2 and 0.7.\\
\hline
IVERM&The number of particles and iterations are 20 and 50, the probability of crossover and mutation are 0.85 and 0.15, the probability of gene exchange in crossover is 0.7, the number of candidate fields is 3.\\
\hline
T-GA&The number of individuals and iterations are 50 and 50, and the probability of crossover and mutation are 0.7 and 0.03.\\
\hline
LB-HGA&The number of individuals and iterations are 40 and 50, the $\lambda_1$, $\lambda_2$ and $\lambda$ are 1.2, 0.8 and 1, the probability of mutation is 0.2.\\
\hline
\end{tabular}
\end{table}
\subsection{Evaluation Results}
In this section, we analyze the performance of the four algorithms according to five evaluation indexes, and give the experimental results and the causes of the results.
\begin{figure}[htbp]
 \center{\includegraphics[width=7.8cm]  {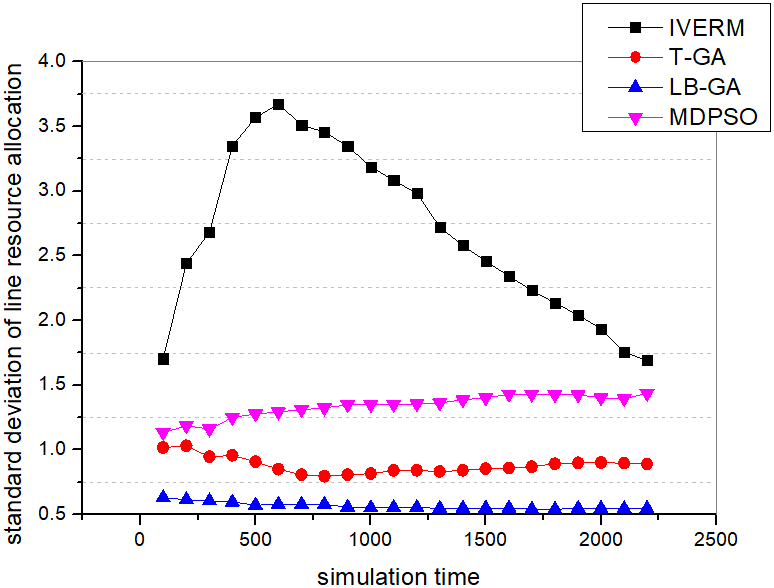}}
 \caption{The diagram of load balancing of the substrate link.}
  \label{fig:8}
\end{figure}

Figure.~\ref{fig:8} uses the standard deviation of resource allocation of the substrate network link as the measurement method of link load balancing. As can be seen from the figure,  LB-HGA algorithm performs best. This is because although the four algorithms all use the shortest path algorithm to map the virtual link, the LB-HGA algorithm considers the link load balancing.
\begin{figure}[htbp]
 \center{\includegraphics[width=7.8cm]  {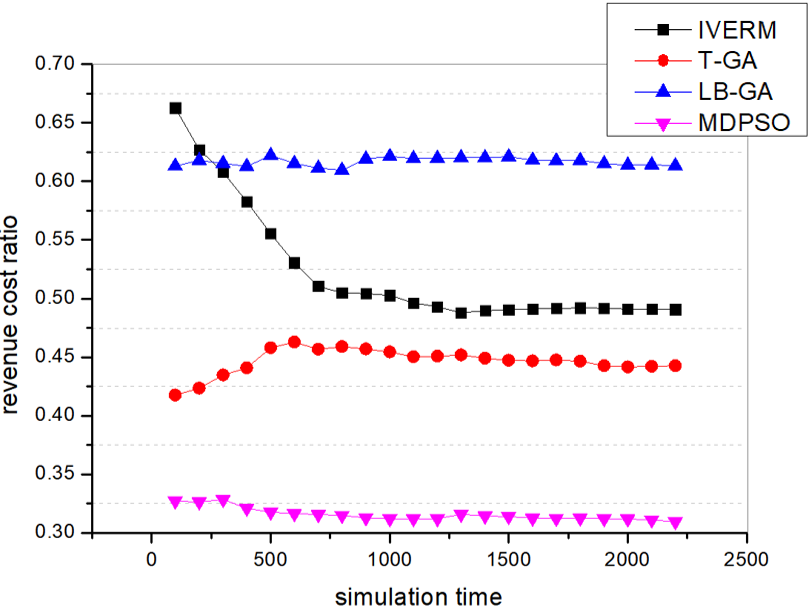}}
 \caption{The diagram of revenue cost ratio.}
 \label{fig:9}
\end{figure}

Figure.~\ref{fig:9} uses revenue cost ratio to compare the resource allocation efficiency of the algorithm. As can be seen from the figure, LB-HGA algorithm performs best. This is because LB-HGA algorithm will obtain the best solution based on fitness, which takes into account price and resource consumption, so the benefit-cost ratio performs well.
\begin{figure}[htbp]
 \center{\includegraphics[width=7.8cm]  {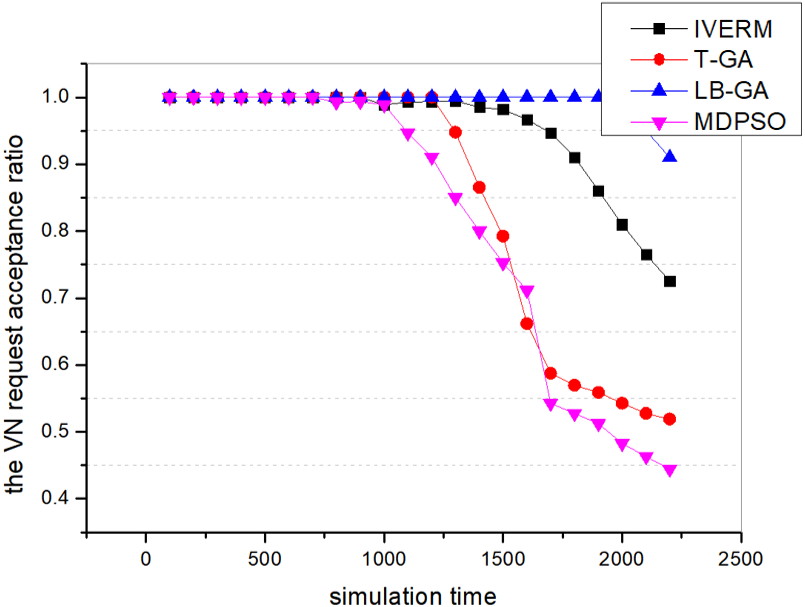}}
 \caption{The diagram of the VN request acceptance ratio.}
 \label{fig:10}
\end{figure}

As can be seen from Fig.~\ref{fig:10}, LB-HGA algorithm performs best in the acceptance rate of virtual network requests. This is because LB-HGA has added the preliminary evaluation of the substrate link resources into the shortest path algorithm, so that the algorithm can bypass the substrate link with insufficient resources, which can avoid most mapping failures. However, the other three algorithms did not clearly consider resource constraints in the link mapping stage, nor did they have a good re-mapping method, so the acceptance rate was poor.
\begin{figure}[htbp]
 \center{\includegraphics[width=7.8cm]  {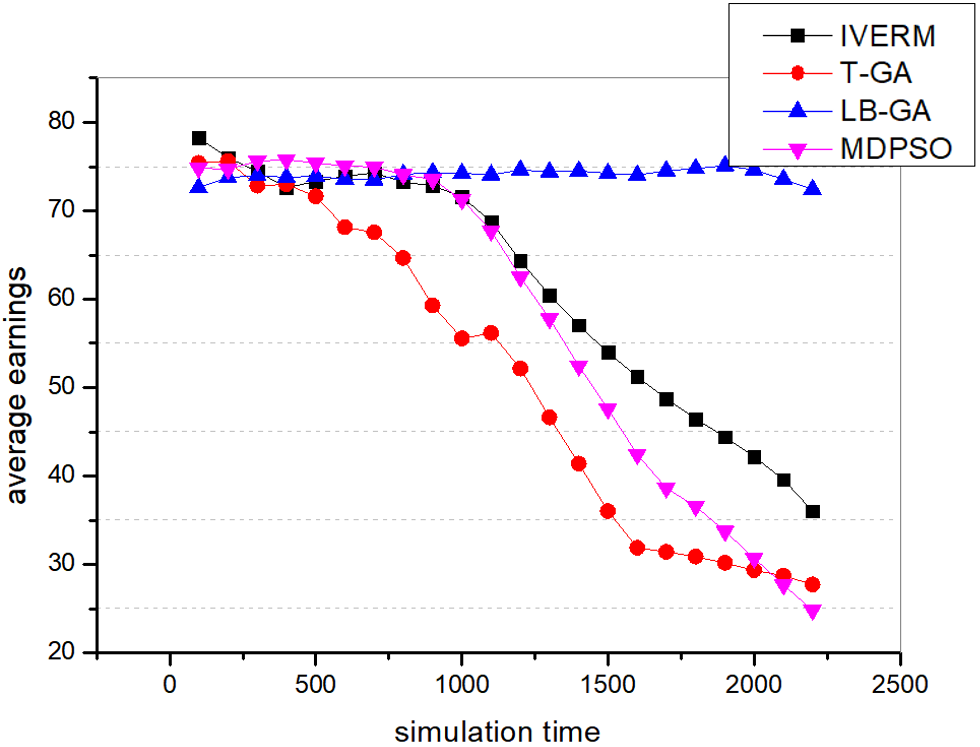}}
 \caption{The diagram of mapping average earnings.}
 \label{fig:11}
\end{figure}

As can be seen from Fig.~\ref{fig:11}, in the early stage when resources are relatively sufficient, the mapping revenue of LB-HGA algorithm is stable, and in the later stage, the revenue will slight decline due to insufficient resources. However, even at an early stage with sufficient resources, the revenue of the other three algorithms is reduced by mapping failures. This can reflect the good performance of LB-HGA algorithm from the side.
\begin{figure}[htbp]
 \center{\includegraphics[width=7.8cm]  {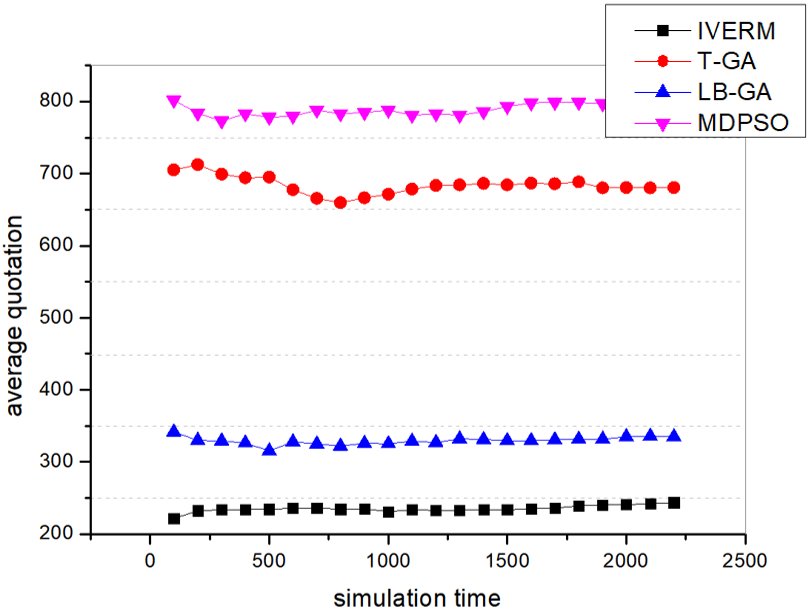}}
 \caption{The diagram of mapping average quotation.}
  \label{fig:12}
\end{figure}

Figure.~\ref{fig:12} uses the product of the resource unit price and the required resource as a measure of the mapping scheme quotation. As can be seen in the figure, the performance of LB-HGA algorithm is second only to IVERM algorithm. This is because because LB-HGA algorithm increased the consideration of load balancing, so the quotation was slightly higher than the IVERM algorithm that gave priority to single domain mapping. However, our algorithm is more stable, which means that our algorithm can get better results with less leeway within the same number of iterations.
\begin{figure}[htbp]
 \center{\includegraphics[width=7.8cm]  {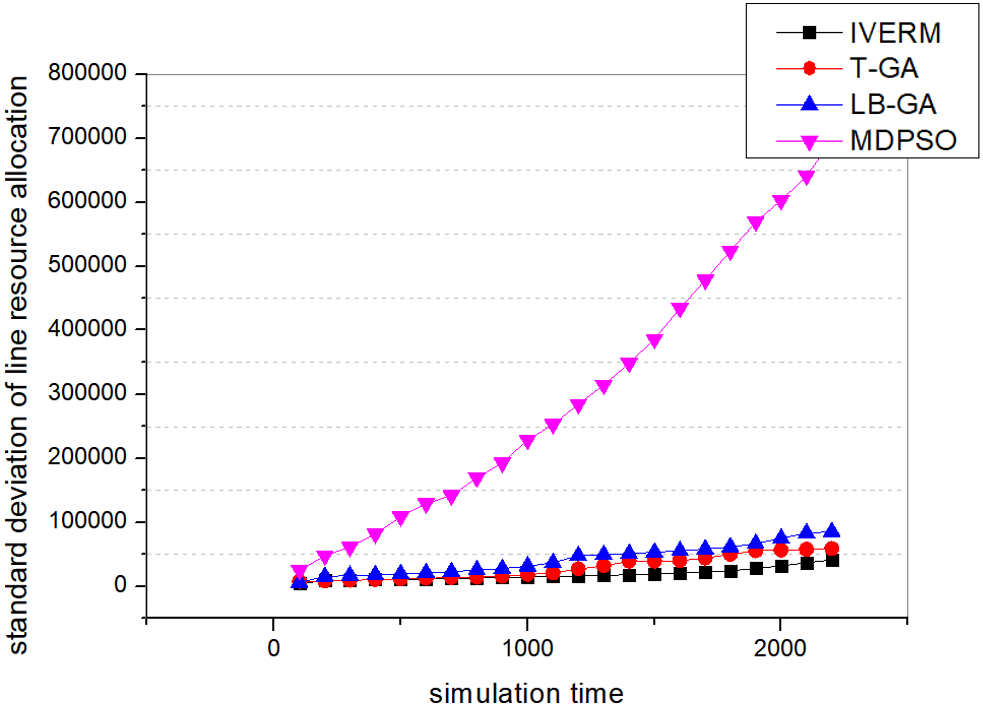}}
 \caption{The diagram of the total running time.}
  \label{fig:13}
\end{figure}

As can be seen from Fig.~\ref{fig:13}, the total running time of IVERM, T-GA, and LB-HGA algorithms are all low and not significantly different. This shows that even if LB-HGA algorithm adds a variety of strategies to ensure the performance of the algorithm, the running time does not increase significantly.
\begin{figure}[htbp]
 \center{\includegraphics[width=7.4cm]  {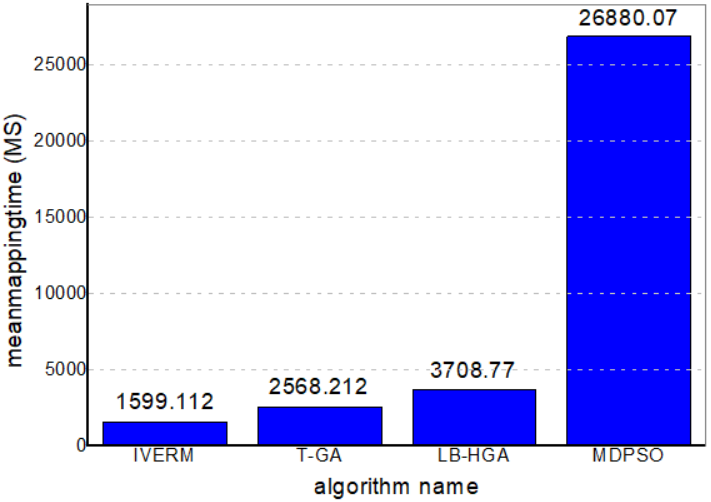}}
 \caption{The diagram of the average total running time.}
  \label{fig:14}
\end{figure}

Figure.~\ref{fig:14} shows the average running time of the four algorithms mapping a virtual network. It can be seen that the running time of LB-HGA algorithm is slightly higher than that of IVERM and T-GA algorithm. This is because the LB-HGA algorithm will re-mapping when the link map fails to improve the VN request acceptance ratio, but this also leads to an increase in the running time. But we use inexact resource constraints to replace precise resource constraints in algorithm iteration, which has reduced the running time as much as possible, making it not much different from other algorithms.
\section{Conclusion}
Heuristic algorithms are suitable for solving NP-hard problems, so they are widely used to solve VNE problems. However, in solving the VNE problem, there are some unresolved problems in the existing work. For example, VNE method based on genetic algorithms usually uses the traditional design method with large randomness, which usually leads to the instability of the quality of the algorithms' results. It is a problem worthy of attention in the Internet of Things environment that requires high network stability and algorithm reliability. In addition, the traditional algorithm's dependence on experience reduces its usefulness, and its low flexibility makes it unable to adapt to increasingly complex network environments. In this paper, the operational optimization of the genetic algorithm is discussed. As a result, the calculation method of crossover probability in three cases is given, as well as the gene scoring strategy for selecting mutated genes. The purpose is to accelerate the convergence speed and make the algorithm more flexible to adapt to different simulation environments. In addition, taking into account different link mapping methods, we analyze the resource constraints and the use of the shortest path algorithm, and we design a link mapping strategy enforcing load balancing. In addition, this strategy improves the accuracy of fitness estimation while improving the acceptance rate by avoiding links with insufficient resources. Simulation results show that our algorithm performs best in link load balance, mapping revenue-cost ratio and VNR acceptance rate, and performs well in mapping average quotation and algorithm running time. In addition, compared with other algorithms, LB-HGA algorithm is significantly more stable and can perform well even in the later stage of the experiment.

In the future work, we will consider better neural network design approaches and hybrid strategies for multiple intelligent algorithms, and we will consider information security in our algorithm. In addition, we intend to study machine learning based algorithms \cite{Jiang2017Machine,Yao2018A} to address the issues of computer networks.
\begin{acknowledgements}
This work is supported by "the Fundamental Research Funds for the Central Universities" of China University of Petroleum (East China) (Grant No. 18CX02139A), the Shandong Provincial Natural Science Foundation, China (Grant No. ZR2014FQ018), and the Demonstration and Verification Platform of Network Resource Management and Control Technology (Grant No. 05N19070040). The authors also gratefully acknowledge the helpful comments and suggestions of the reviewers, which have improved the presentation.
\end{acknowledgements}

\bibliographystyle{ieeetr}
\bibliography{references}

\end{document}